\documentstyle[epsfig]{elsart}

\def\ageq{\vcenter{\vbox{\hbox{$\buildrel > \over \sim$}}}}

\begin{document}
\begin{frontmatter}
\title{Exclusive photo- and electroproduction of mesons \\
at large momentum transfer\thanksref{talk}}
\thanks[talk]{Talk given by W. Schweiger at the 2$^{\rm nd}$ ELFE Workshop, St.
Malo, Sep. 1996.}
\author{G. Folberth, R. Rossmann and W. Schweiger}
\address{Institute of Theoretical Physics, University of Graz, \\
A-8010 Graz, Universit\"atsplatz 5, Austria}

\begin{abstract}
We analyze the reactions $\gamma^{(\ast)}\, {\rm p}\, \rightarrow \, {\rm M}\, {\rm
B}$, with ${\rm M}\, {\rm B}$ being either ${\rm K}^+\, \Lambda$,  ${\rm K}^+\,
\Sigma^0$, or $\pi^+\, {\rm n}$, within perturbative QCD, allowing for diquarks as
quasi-elementary constituents of baryons. The diquark-model parameters and the
quark-diquark distribution amplitudes of the baryons are taken from previous
investigations of electromagnetic baryon form factors and Compton-scattering off
protons. Reasonable agreement with the few existing data at large momentum transfer
is achieved if the asymptotic form ($\propto x (1-x)$) is chosen for the meson
distribution amplitudes.
\end{abstract}
\end{frontmatter}

Our investigation of exclusive photo- and electroproduction of mesons continues a
systematic study of hard exclusive reactions \cite{Ja93,Kro93,Kro96}  within a model
which is based on perturbative QCD, in  which baryons, however, are treated as
quark-diquark systems. 
This model has been successfully applied to the description of 
baryon form factors in the space- \cite{Ja93} and time-like region \cite{Kro93}, real
and virtual Compton scattering \cite{Kro96}, two-photon annihilation into
proton-antiproton \cite{Kro93}. and the charmonium decay $\eta_{{\rm c}} \rightarrow
{\rm p}\, \bar{{\rm p}}$ \cite{Kro93}. Like the usual hard-scattering approach (HSA)
\cite{BL89} the diquark-model relies on factorization of short- and long-distance
dynamics; a hadronic amplitude is expressed as a convolution of a hard-scattering
amplitude $\widehat{T}$, calculable within perturbative QCD, with distribution
amplitudes (DAs) $\phi$ which contain the (non-perturbative) bound-state dynamics of
the hadronic constituents. The introduction of diquarks does not only simplify
computations, it is rather motivated by the requirement to extend the HSA from
(asymptotically) large down to intermediate momentum transfers ($p_{\perp}^2 \,
\ageq \, 4 \hbox{GeV}^2$). This is the momentum-transfer region where some
experimental data already exist, but where still persisting non-perturbative
effects, in particular strong correlations in baryon wave functions, prevent the
pure quark HSA to become fully operational. Diquarks may be considered as an
effective way to cope with such effects. 

The hard-scattering amplitude $\widehat{T}$ is process dependent and represents the
scattering of the hadronic constituents in collinear approximation. It consists of
all posssible tree graphs obtained by replacing the external hadrons by their valence
Fock-state (in our model a quark-diquark state in case of a baryon) and by
distributing the total momentum transfer over all the hadronic constituents by means
of gluon exchange. The model, as applied in Refs.~\cite{Ja93,Kro93,Kro96}, comprises
scalar (S) as well as axial-vector (V) diquarks. V-diquarks are important, if one
wants to describe spin observables which require the flip of baryonic helicities.
The Feynman rules for electromagnetically interacting diquarks are just those of
standard quantum electrodynamics. Feynman rules for strongly interacting diquarks are
obtained by replacing the electric charge $e$ by the strong coupling constant $g_{\rm
s}$ times the Gell-Mann colour matrix $t^a$. The composite nature of diquarks is
taken into account by multiplying each of the Feynman diagrams with diquark form
factors which depend on the kind of the diquark (S or V) and the number of gauge
bosons coupling to the diquark.  The form factors are  parameterized by  multipole
functions  with the power chosen in such a way that in the limit $p_{\perp}
\rightarrow \infty$ the scaling behaviour of the pure quark HSA is recovered.

The process independent DAs are, roughly speaking, valence Fock-state wave functions
integrated over the transverse momentum (up to a factorization scale
$\tilde{p}_{\perp}$ which depends on the momentum transfer $p_{\perp}$).  In
\cite{Ja93,Kro93,Kro96} a quark-diquark DA of the form ($c_1 = c_2 = 0$ for S
diquarks, x ..... longitudinal momentum fraction carried by the quark)
$$
\phi_{{\rm D}}^{{\rm B}}  \propto x (1-x)^3 (1 + c_1 x + c_2 x^2)
\exp \left[ - b^2 \left( \frac{m_{{\rm q}}^2}{x} + \frac{m_{{\rm D}}^2 }{
(1-x)} \right) \right] \, , \quad {\rm D = S, V} \, ,
$$
in connection with an SU(6)-like spin-flavour wave function, turned out to be quite
appropriate for octet baryons B. In order to check the dependence on the choice of
the meson DAs we employ two qualitatively different forms. On the one hand the
asymptotic DA
$$
\phi_{\rm asy} \propto x (1-x) \, , 
$$
which solves the $\tilde{p}_{\perp}$ evolution equation for $\phi(x,
\tilde{p}_{\perp})$ in the limit $\tilde{p}_{\perp} \rightarrow \infty$, and on the
other hand the DAs
$$
\phi_{\rm CZ}^{\pi} \propto \phi_{\rm asy} (2 x -1)^2 \, , \quad
\phi_{\rm CZ}^{\rm K} \propto \phi_{\rm asy} [0.08 + 0.6 (2 x -1)^2 + 0.25 (2 x
-1)^3] \, ,
$$
which have been proposed in Ref.~\cite{CZ84} on the basis of QCD sum rules. The
\lq\lq normalization\rq\rq of the meson DAs is determined by the experimental decay 
constants for the weak $\pi$, K $\rightarrow \, \mu \, \nu_{\mu}$ decays. 

With these DAs and the diquark-model parameters of Ref.~\cite{Ja93} one obtains the
results depicted in Figs.~1 and 2. The three production channels we are considering
differ qualitatively in the sense that ${\rm K}^+ \, \Lambda$ is solely produced via
the S$_{\rm [u,d]}$ diquark, ${\rm K}^+ \, \Sigma^0$ via the V$_{\rm \{u,d\}}$
diquark, and $\pi^+ \, {\rm n}$ via S and V diquarks. An important consequence of
this observation is that spin observables which require the flip of the
baryonic helicity are predicted to vanish for the \mbox{${\rm K}^+$-$\Lambda$} final
state (e.g., the polarization $P$ of the outgoing $\Lambda$). For the other two
channels helicity flips may, of course, take place by means of the V diquark.  In
all three channels the photoproduction data are better reproduced with the asymptotic
meson DA $\phi_{\rm asy}$, whereas $\phi_{\rm CZ}$ tends to exhibit an overshooting
of the data. This finding is in line with the conclusions drawn from the
investigation of the pion-photon transition form factor \cite{JKR96}. There, strongly
end-point concentrated DAs, like $\phi_{\rm CZ}^{\pi}$ are also ruled out by the data.
Photoproduction calculations within the pure quark HSA \cite{FHZ91} yield less
satisfactory results. The difference between the two meson DAs becomes even more
obvious, if one considers spin-dependent quantities. As an example we have plotted in
Fig.~1 (right) the cross section ratios $d\sigma_{\rm L}/ d\sigma_{\rm T}$ for two of
the pieces entering the cross section for electroproduction of the ${\rm K}^+$-$\Lambda$ 
state. $d\sigma_{\rm L(T)}/dt$ can be understood as the cross section for
photoproduction with a longitudinally (transversally) polarized virtual photon. For
the photon virtuality $Q^2 \rightarrow 0$ the cross section $d\sigma_{\rm T}/dt$ goes
over in the usual photoproduction cross section. A more detailed discussion of
$\gamma\, {\rm p}\, \rightarrow \, {\rm K}^+\, {\Lambda}$ (and also $\gamma\, {\rm
p}\, \rightarrow \, {{\rm K}^{\ast}}^+\, {\Lambda}$) can be found in
Ref.~\cite{KSPS96}. This paper contains also analytic expressions for the
photoproduction amplitudes and  describes in detail our calculational techniques. 
\begin{figure}
\begin{center}
\epsfig{file=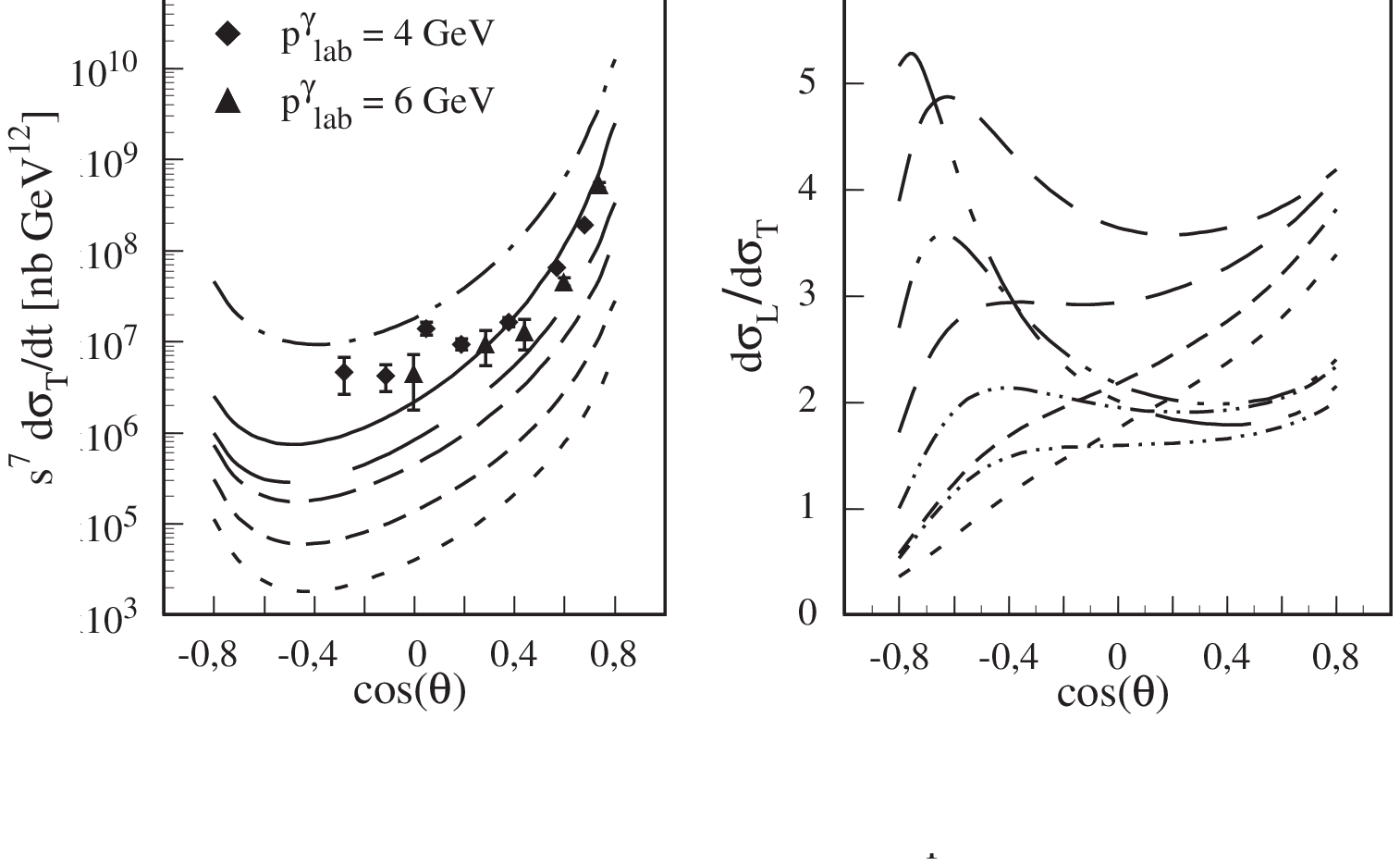, height=6cm}
\end{center}
\caption[]{Diquark-model predictions for photo- and electroprduction of the
K$^+$-$\Lambda$ final state; results are shown for $p^{\gamma}_{\rm lab}$ = 6~GeV
and photon virtualities $Q^2$ of 0 (photoproduction), 0.5, 1, 2, and 3 GeV$^2$. The
dashes become  shorter with increasing $Q^2$.
\\
{\em Left figure:} The cross section $d\sigma_{\rm T}/dt$  (solid and dashed lines),
obtained with the asymptotic Kaon DA $\phi_{\rm asy}$. The dash-dotted curve is the
photoproduction result obtained with the Chernyak-Zhitnitsky DA $\phi_{\rm CZ}^{\rm
K}$ for the Kaon. The experimental points are photoproduction data taken from Anderson
et al.
\cite{An76}. 
\\ 
{\em Right Figure:} Cross section ratio $d\sigma_{\rm L}/d\sigma_{\rm T}$ for
$\phi_{\rm asy}$ (dashed lines) and $\phi_{\rm CZ}^{\rm K}$ (dash-dotted lines).}
\end{figure}
\begin{figure}
\begin{center}
\epsfig{file=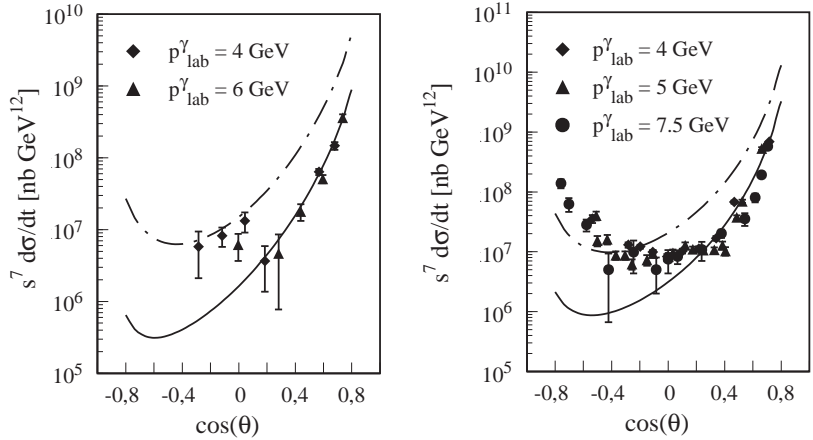, height=6cm}
\end{center}
\caption[]{Diquark-model predictions for photoproduction of the
K$^+$-$\Sigma^0$ ({\em left figure}) and the $\pi^+$-n ({\em right figure}) final
states; results are shown for 
\mbox{$p^{\gamma}_{\rm lab}$ = 6~GeV.}  The solid lines
have been obtained with $\phi_{\rm asy}$, the dash-dotted lines with $\phi_{\rm CZ}^{\rm
K}$ and $\phi_{\rm CZ}^{\pi}$, repectively. Data are taken from Anderson et al.
\cite{An76}.}
\end{figure}
\newpage

To summarize: The predictions of the diquark model for $\gamma\, {\rm p}\, 
\rightarrow$ ${\rm K}^+\, \Lambda$,  ${\rm K}^+\, \Sigma^0$, and $\pi^+\, {\rm n}$ 
look rather promising if the asymptotic form ($\propto x (1-x)$) is taken for the
meson DAs. To the best of our knowledge the diquark model is, as yet, the only
constituent scattering model which is able to account for the large-$p_{\perp}$
photoproduction data. With respect to future experiments it would, of course, be
desirable to have more and better large momentum-transfer data. Polarization
measurements of the recoiling particle could help to decide, whether the
perturbative regime has been reached already, or whether non-perturbative effects
(different from diquarks) are still at work. Polarization observables, in general,
and also electroproduction data at large $p_{\perp}$ could put additional
constraints on the form of the hadron DAs.


\vspace{-0.5 cm}


\begin{thebibliography}{99}

\vspace{-0.5 cm}

\bibitem{Ja93}R. Jakob et al., Z. Phys. {\bf A347} (1993) 109
\bibitem{Kro93}P. Kroll et al., Phys. Lett. {\bf B316} (1993) 546
\bibitem{Kro96}P. Kroll, M. Sch\"urmann, and P. A. M. Guichon, Nucl. Phys. {\bf
A598} (1996) 435
\bibitem{BL89}See, e.g., S. J. Brodsky and G. P. Lepage, in {\it Perturbative Quantum
Chromodynamics}, edt. by A. H. Mueller (World Scientific, Singapore, 1989)
\bibitem{CZ84}V. L. Chernyak and A. R. Zhitnitsky, Phys. Rep. {\bf 112} (1984) 173
\bibitem{An76}R. L. Anderson et al., Phys. Rev. {\bf D14} (1976) 679
\bibitem{JKR96}R. Jakob, P. Kroll, and M. Raulfs, J. Phys. {\bf G22} (1996) 45
\bibitem{FHZ91}G. R. Farrar, K. Huleihel, and H. Zhang, Nucl. Phys. {\bf B349} (1991)
655
\bibitem{KSPS96} P. Kroll et al., preprint UNIGRAZ-UTP 15-04-96 (hep-ph/9604353)
\end{thebibliography}
\end{document}